\documentclass[journal]{IEEEtran}

\usepackage{amsmath,amsfonts,amssymb,amsmath,latexsym,texdraw}

\newtheorem{theorem}{Theorem}
\newtheorem{lemma}[theorem]{Lemma}
\newtheorem{corollary}[theorem]{Corollary}

\begin{document}
%
\title{Weight Distributions of Hamming Codes}
%
%
\author{Dae San Kim,~\IEEEmembership{Member,~IEEE}
\thanks{This work was supported by grant No. R01-2007-000-11176-0 from the Basic Research Program of the Korea Science and Engineering Foundation.}
\thanks{The author is with the Department of Mathematics, Sogang University, Seoul 121-742, Korea(e-mail: dskim@sogang.ac.kr). }
        }
\maketitle

\begin{abstract}
We derive a recursive formula determining the weight distribution of
the $[n=(q^m-1)/(q-1),~n-m,~3]$ Hamming code $H(m,~q)$, when
$(m,~q-1)=1$. Here $q$ is a prime power. The proof is based on
Moisio's idea of using Pless power moment identity together with
exponential sum techniques.
\end{abstract}



%
\IEEEpeerreviewmaketitle

\section{Introduction}

The Hamming code is probably the first one that someone encounters
when he is taking a beginning course in coding theory. The $q$-ary
Hamming code $H(m,~q)$ is an $[n=(q^m-1)/(q-1),~n-m,~3]$ code which
is a single-error-correcting perfect code. From now on, $q$ will
indicate a prime power unless otherwise stated. Also, we assume
$m>1$.

In \cite{M01}, Moisio discovered a handful of new power moments of
Kloosterman sums over $\mathbb{F}_q$, with $q=2^r$. This was done,
via Pless power moment identity, by connecting moments of
Kloosterman sums and frequencies of weights in the binary Zetterberg
code of length $q+1$, which were known by the work of Schoof and van
der Vlugt in \cite{SV}. Some new moments of Kloosterman sums were
also found over $\mathbb{F}_q$, with $q=3^r$
(\cite{M02},\cite{GSV}).

In this correspondence, we adopt Moisio's idea of utilizing Pless
power moment identity and exponential sum techniques and prove the
following theorem giving the weight distribution of $H(m,q)$, for
$(m,q-1)=1$.

\begin{theorem}\label{mt}
    Let $\{C_h\}_{h=0}^{n}(n=(q^m-1)/(q-1))$ denote the weight
    distribution of the $q$-ary Hamming code $H(m,~q)$, with
    $(m,~q-1)=1$. Then, for $h$ with $1\leq h \leq n$,
    \begin{align*}
        h!C_h=&(-1)^hq^{m(h-1)}(q^m-1)\\
              &+\sum_{i=0}^{h-1}(-1)^{h+i-1}C_i\sum_{t=i}^{h}t!S(h,t)q^{h-t}(q-1)^{t-i}(^{n-i}_{n-t}),
    \end{align*}
    where $S(h,t)$ denotes the Stirling number of the second kind
    defined by
    \begin{equation}\label{snsk}
        S(h,t)=\frac{1}{t!}\sum_{j=0}^{t}(-1)^{t-j}(^t_j)j^h.
    \end{equation}
\end{theorem}

$C_0=1$, and it is easy to check that $C_1=C_2=0$, as it should be.
A few next values of $C_h$'s were obtained, with the help of
\textit{Mathematica}, from the above formula.

\begin{corollary}
    Let $\{C_h\}_{h=0}^n (n=(q^m-1)/(q-1))$ denote the weight distribution of the
    $q$-ary Hamming code $H(m,q)$, with $(m,q-1)=1$. Then
    \begin{align*}
        C_3=&\frac{1}{3!}(q^m-1)(-q+q^m),\\
        C_4=&\frac{1}{4!}(q^m-1)(-6q+5q^2+6q^m+q^{2m}-6q^{1+m}),\\
        C_5=&\frac{1}{5!}(q^m-1)(-36q+54q^2-26q^3+36q^m+6q^{2m}\\
            &+q^{3m}-60q^{1+m}+35q^{2+m}-10q^{1+2m}),\\
        C_6=&\frac{1}{6!}(q^m-1)(-240q+500q^2-450q^3+154q^4\\
            &+240q^m+20q^{2m}+10q^{3m}+q^{4m}-520q^{1+m}\\
            &+85q^{2(1+m)}+550q^{2+m}-225q^{3+m}-110q^{1+2m}\\
            &-15q^{1+3m}),\\
        C_7=&\frac{1}{7!}(q^m-1)(-1800q+4710q^2-6035q^3+3940q^4\\
            &-1044q^5+1800q^m-90q^{2m}+85q^{3m}+15q^{4m}\\
            &+q^{5m}-4620q^{1+m}+1505q^{2(1+m)}+6755q^{2+m}\\
            &-5215q^{3+m}+1624q^{4+m}-805q^{1+2m}-735q^{3+2m}\\
            &-245q^{1+3m}+175q^{2+3m}-21q^{1+4m}),\\
        C_8=&\frac{1}{8!}(q^m-1)(-15120q+47124q^2-77196q^3+72779q^4\\
            &-37240q^5+8028q^6+15120q^m-3276q^{2m}+840q^{3m}\\
            &+175q^{4m}+21q^{5m}+q^{6m}-43848q^{1+m}\\
            &+17934q^{2(1+m)}-1960q^{3(1+m)}+ 79632q^{2+m}\\
            &+6769q^{2(2+m)}-87808q^{3+m}+ 52661q^{4+m}\\
            &-13132q^{5+m}-3276q^{1+2m}-19236q^{3+2m}\\
            &-3080q^{1+3m}+4270q^{2+3m}-476q^{1+4m}+322q^{2+4m}\\
            &-28q^{1+5m}),\\
        C_9=&\frac{1}{9!}(q^m-1)(-141120q+507024q^2-1002736q^3\\
            &+1221444q^4-910644q^5+382088q^6-69264q^7\\
            &+141120q^m-57456q^{2m}+10864q^{3m}+1960q^{4m}\\
            &+322q^{5m}+28q^{6m}+q^{7m}-449568q^{1+m}\\
            &+165396q^{2(1+m)}-67116q^{3(1+m)}+ 957936q^{2+m}\\
            &+246624q^{2(2+m)}-1349404q^{3+m}+1175874q^{4+m}\\
            &-571116q^{5+m}+118124q^{6+m}+33936q^{1+2m}\\
            &-332584q^{3+2m}-67284q^{5+2m}-39396q^{1+3m}\\
            &+74844q^{2+3m}+22449q^{4+3m}-7812q^{1+4m}\\
            &+10332q^{2+4m}-4536q^{3+4m}-840q^{1+5m}\\
            &+546q^{2+5m}-36q^{1+6m}),
    \end{align*}
    and
    \begin{align*}
        C_{10}=&\frac{1}{10!}(q^m-1)(-1451520q+5880384q^2-13550832q^3\\
               &+20090832q^4-19485852q^5+11984244q^6\\
               &-4251240q^7+663696q^8+ 1451520q^m-893376q^{2m}\\
               &+174384q^{3m}+ 21504q^{4m}+4536q^{5m}+546q^{6m}\\
               &+36q^{7m}+q^{8m}-4987008q^{1+m}+ 857520q^{2(1+m)}\\
               &-1569540q^{3(1+m)}+63273q^{4(1+m)}+12035088q^{2+m}\\
               &+5797770q^{2(2+m)}-20393616q^{3+m}+723680q^{2(3+m)}\\
               &+23050848q^{4+m}-16423398q^{5+m}+6661236q^{6+m}\\
               &-1172700q^{7+m}+1341360q^{1+2m}-4686480q^{3+2m}\\
               &-3264780q^{5+2m}-576240q^{1+3m}+1233960q^{2+3m}\\
               &+1030260q^{4+3m}-269325q^{5+3m}-117012q^{1+4m}\\
               &+227808q^{2+4m}-196392q^{3+4m}-17430q^{1+5m}\\
               &+22260 q^{2+5m}-9450q^{3+5m}-1380q^{1+6m}\\
               &+870q^{2+6m}-45q^{1+7m}).
    \end{align*}
\end{corollary}

The Hamming code was discovered by Hamming in late 1940's. So it is
surprising that there are no such recursive formulas determining the
weight distributions of the Hamming codes in the nonbinary cases. In
the binary case, we have the following well known formula which
follows from elementary combinatorial reasoning(\cite[p. 129]{MS}).

\begin{theorem}
    Let $\{C_h\}_{h=0}^{n}(n=(2^m-1))$ denote the weight distribution of the binary Hamming
    code $H(m,2)$. Then the weight distribution satisfies the following recurrence relation:
    \begin{align*}
        &C_0=1, ~C_1=0,\\
        &(i+1)C_{i+1}+C_i+(n-i+1)C_{i-1}=(^n_i)~~(i\geq 1).
    \end{align*}
\end{theorem}

It is known \cite{PHB} that, when $(m,q-1)=1$, $H(m,q)$ is a cyclic
code.

\begin{theorem}
    Let $n=(q^m-1)/(q-1)$, where $(m,q-1)=1$.
    Let $\gamma$ be a primitive element of $\mathbb{F}_{q^m}$.
    Then the cyclic code of length $n$ with the defining zero $\gamma^{q-1}$
    is equivalent to the $q$-ary Hamming code $H(m,q)$.
\end{theorem}

In our discussion below, we will assume that $(m,q-1)=1$, so that
$H(m,q)$ is a cyclic code with the defining zero $\gamma^{q-1}$,
where $\gamma$ is a primitive element of $\mathbb{F}_{q^m}$.

\section{Preliminaries}

Let $q=p^r$ be a prime power. Then we will use the following
notations throughout this correspondence.

\begin{align*}
    tr(x)=&x+x^p+\cdots +x^{p^{r-1}}\\
          &the ~trace ~function ~\mathbb{F}_q\rightarrow \mathbb{F}_p,\\
    Tr(x)=&x+x^q+\cdots +x^{q^{m-1}}\\
          &the ~trace ~function ~\mathbb{F}_{q^m}\rightarrow \mathbb{F}_q,\\
    \lambda(x)=&e^{\frac{2\pi i}{p}tr(x)}\\
               &the ~canonical ~additive ~character ~of ~\mathbb{F}_q,\\
    \lambda_m(x)=&\lambda(Tr(x))\\
                 &the ~canonical ~additive ~character ~of ~\mathbb{F}_{q^m}.
\end{align*}

The following lemma is well known.

\begin{lemma}\label{ac}
    For any $\alpha \in \mathbb{F}_q$,
    \begin{equation*}
        \sum_{x\in \mathbb{F}_q}\lambda(\alpha
        x)=\left\{\begin{array}{ll} q,&\alpha=0,\\0,&\alpha \neq
        0.\end{array}\right.
    \end{equation*}
\end{lemma}

For a positive integer $s$, the multiple Kloosterman sum
$K_s(\alpha)~(\alpha \in \mathbb{F}_q^*)$, is defined by

\begin{equation*}
    K_s(\alpha)=\sum_{x_1,\cdots ,x_s \in \mathbb{F}_q^*}\lambda(x_1+\cdots
    +x_s+\alpha x_1^{-1}\cdots x_s^{-1}).
\end{equation*}

The following result follows immediately from Lemma \ref{ac}.

\begin{lemma}\label{k}
    For an integer $s>1$,
    \begin{equation*}
        \sum_{\alpha \in \mathbb{F}_q^*}K_{s-1}(\alpha)=(-1)^s.
    \end{equation*}
\end{lemma}

\begin{proof}
    $\sum_{\alpha \in \mathbb{F}_q^*}K_{s-1}(\alpha)=(\sum_{x \in
    \mathbb{F}_q^*}\lambda(x))^s$.
\end{proof}

The following lemma is immediate.

\begin{lemma}\label{b}
    Let $(m,q-1)=1$. Then the following map is a bijection.
    \begin{equation*}
        \alpha \mapsto \alpha^m: \mathbb{F}_q^* \rightarrow \mathbb{F}_q^*.
    \end{equation*}
\end{lemma}

\begin{theorem}[Thm. 3 of \cite{M03}]\label{tofm}
    For any $\alpha \in \mathbb{F}_{q^m}^*$,
    \begin{equation*}
        \sum_{x \in \mathbb{F}_{q^m}^*}\lambda_m(\alpha
        x^{q-1})=(-1)^{m-1}(q-1)K_{m-1}(N(\alpha)),
    \end{equation*}
    where $N$ denotes the norm map $N: \mathbb{F}_{q^m}^* \rightarrow \mathbb{F}_q^*$,
    defined by $N(\alpha)=\alpha^n$, with $n=(q^m-1)/(q-1)$.
\end{theorem}

The following theorem is due to Delsarte(\cite[P. 208]{MS}).

\begin{theorem}[Delsarte]\label{tofd}
    Let $B$ be a linear code of length $n$ over $\mathbb{F}_{q^m}$. Then
    \begin{equation*}
        (B|_{\mathbb{F}_q})^\bot=Tr(B^\bot).
    \end{equation*}
\end{theorem}

The following is a special case of the result stated in \cite[Thm.
4.2]{HT}, although only the binary case is mentioned there. In fact,
using Theorem \ref{tofd} above, this can be proved in exactly the
same manner as described immediately after the proof of Theorem 4.2
in \cite{HT}.

\begin{theorem}\label{dh}
    The dual $H(m,q)^\bot$ of $H(m,q)$ is given by
    \begin{align*}
        H(m,&q)^\bot\\
        =\{&c(a)=(Tr(a),Tr(a \gamma^{(q-1)}),\cdots ,Tr(a \gamma^{(n-1)(q-1)}))\\
           &|a \in \mathbb{F}_{q^m}\}.
    \end{align*}
\end{theorem}

\begin{lemma}\label{i}
    The map $a\mapsto c(a): \mathbb{F}_{q^m}\rightarrow H(m,q)^
    \bot$ is an isomorphism of $\mathbb{F}_q$-vector spaces.
\end{lemma}

\begin{proof}
    The map is $\mathbb{F}_q$-linear, surjective and
    $dim_{\mathbb{F}_q}\mathbb{F}_{q^m}=dim_{\mathbb{F}_q}H(m,q)^\bot$.
\end{proof}

Our recursive formula in Theorem \ref{mt} will be a consequence of
the application of Pless power moment identity(\cite{PHB}), which is
equivalent to MacWilliams identity.

\begin{theorem}[Pless power moment identity]\label{ppmi}
    Let $B$ be an $q$-ary $[n,k]$ code,
    and let $B_i$(resp. $B_i^\bot$) denote the number of codewords of weight $i$ in $B$(resp.
    in $B^\bot$). Then, for $h=0,1,2,\cdots,$
    \begin{align}\label{eofp}
        \sum_{i=0}^{n}i^hB_i=\sum_{i=0}^{min\{n,h\}}(-1)^iB_i^\bot\sum_{t=i}^{h}t!S(h,t)q^{k-t}(q-1)^{t-i}(^{n-i}_{n-t}),
    \end{align}
    where $S(h,t)$ denotes the Stirling number of the second kind
    defined by (\ref{snsk}).
\end{theorem}

\section{Proof of Theorem 1}

Let $h$ be an integer with $1\leq h \leq n$. Observe that the weight
of the codeword $c(a)$ in Theorem \ref{dh} can be expressed as

\begin{align}\label{poft}
    \begin{split}
    w(c(a))=&\sum_{i=0}^{n-1}(1-q^{-1}\sum_{\alpha \in \mathbb{F}_q}\lambda(\alpha
    Tr(a\gamma^{i(q-1)})))\\
     &\hspace*{4.6cm}(\text{by Lemma} ~\ref{ac})\\
    =&n-q^{-1}\sum_{\alpha \in \mathbb{F}_q}\sum_{i=0}^{n-1}\lambda_m(\alpha a \gamma^{i(q-1)})\\
    =&n-q^{-1}(q-1)^{-1}\sum_{\alpha \in \mathbb{F}_q}\sum_{x \in \mathbb{F}_{q^m}^*}\lambda_m(\alpha a x^{q-1})\\
    =&n-q^{-1}(q-1)^{-1}(q^m-1)-q^{-1}(q-1)^{-1}\\
     &\times\sum_{\alpha \in \mathbb{F}_q^*}\sum_{x \in \mathbb{F}_{q^m}^*}\lambda_m(\alpha a x^{q-1})\\
    =&n-q^{-1}(q-1)^{-1}(q^m-1)+(-1)^mq^{-1}\\
     &\times\sum_{\alpha \in \mathbb{F}_q^*}K_{m-1}(\alpha^mN(a))\hspace*{.7cm}(\text{by Theorem} ~\ref{tofm})\\
    =&n-q^{-1}(q-1)^{-1}(q^m-1)+(-1)^mq^{-1}\\
     &\times\sum_{\alpha \in \mathbb{F}_q^*}K_{m-1}(\alpha N(a)).\hspace*{1.1cm}(\text{by Lemma} ~\ref{b})\\
    \end{split}
\end{align}

We now apply Pless power moment identity in Theorem \ref{ppmi} with
$B=H(m,q)^\bot$. On one hand, the LHS of (\ref{eofp}) is

\begin{align*}
    \sum_{a \in \mathbb{F}_{q^m}^*}&w(c(a))^h \hspace*{4cm}(\text{by Lemma} ~\ref{i})\\
    =&\sum_{a \in \mathbb{F}_{q^m}^*}(n-q^{-1}(q-1)^{-1}(q^m-1)+(-1)^mq^{-1}\\
     &\times\sum_{\alpha \in \mathbb{F}_q^*}K_{m-1}(\alpha N(a)))^h \hspace*{2.7cm}(\text{by} ~(\ref{poft}))\\
    =&\frac{q^m-1}{q-1}\sum_{a \in \mathbb{F}_q^*}(n-q^{-1}(q-1)^{-1}(q^m-1)+(-1)^mq^{-1}\\
     &\times\sum_{\alpha \in \mathbb{F}_q^*}K_{m-1}(\alpha a))^h\\
    =&(q^m-1)(n-q^{-1}(q-1)^{-1}(q^m-1)+(-1)^mq^{-1}\\
     &\times\sum_{\alpha \in \mathbb{F}_q^*}K_{m-1}(\alpha))^h\\
    =&(q^m-1)(n-q^{-1}(q-1)^{-1}(q^m-1)+q^{-1})^h\\
     &\hspace*{5.5cm}(\text{by Lemma} ~\ref{k})\\
    =&q^{(m-1)h}(q^m-1) ~(as ~n=\frac{q^m-1}{q-1}).
\end{align*}

On the other hand, by separating the term corresponding to $h$ and
noting $S(h,h)=1$, the RHS of (\ref{eofp}) is
\begin{align*}
    &(-1)^hC_hh!q^{m-h}\\
    &+\sum_{i=0}^{h-1}(-1)^iC_i\sum_{t=i}^{h}t!S(h,t)q^{m-t}(q-1)^{t-i}(^{n-i}_{n-t}).
\end{align*}
So
\begin{align}\label{moft}
    \begin{split}
    q^{(m-1)h}&(q^m-1)\\
    =&(-1)^hC_hh!q^{m-h}\\
     &+\sum_{i=0}^{h-1}(-1)^iC_i\sum_{t=i}^{h}t!S(h,t)q^{m-t}(q-1)^{t-i}(^{n-i}_{n-t}).
    \end{split}
\end{align}
 Multiplying both sides of (\ref{moft}) by $(-1)^hq^{h-m}$, we get the desired
 result.$\hspace*{6.5cm}\blacksquare$

%








\end{document}